\documentclass[12pt,preprint]{aastex}
\def\NH{\ifmmode N_{\rm H}\else$N_{\rm H}$\fi}
\def\NHlit{\ifmmode N_{\rm H,lit.}\else$N_{\rm H,lit.}$\fi}
\begin{document}

\title{Multi-wavelength variability of the
magnetar 4U~0142+61}
\author{Martin Durant and Marten H. van Kerkwijk}
\affil{Istituto de Astrof\'izica de Canarias \\
C/ V\'ia L\'actea, s/n \\
E38205 - La Laguna (Tenerife), Spain
\\ ~ \\
Department of Astronomy and Astrophysics, University of
  Toronto\\  60 St. George St, Toronto, ON\\ M5S 3H8, Canada }

\begin{abstract}
We have collected data spanning seven years of observations of the magnetar
4U~0142+61 in the infrared, optical and soft X-rays. These combine our
own observations and analysis of archival 
data. We find that the source is variable in the optical, in contrast
to what had been previously reported, that the K-band flux can vary by
over a magnitude on the time-scale of days, and that the X-ray pulsed flux is
not obviously correlated with either the total X-ray flux or infrared
and optical fluxes. Furthermore, from multi-color photometry of the
source within single nights, we conclude that there are two separate
components to the infrared emission. The overall picture is unclear,
and prompts the need for further, more frequent observations.
\end{abstract}

\section{Introduction}
The Anomalous X-ray Pulsars (AXPs) are a group of about six young,
energetic neutron stars. They are termed {\em anomalous} since their energy
source was initially not known: the rotational spin-down luminosities
are too low and no binary companions are found. They are now modeled as {\em
  magnetars} along with the Soft Gamma-ray Repeaters. Magnetars are
neutron stars with external magnetic fields of the order $10^{15}$\,G
and even stronger internal fields. It is this magnetic field which
acts as an energy reservoir and powers the observed emission as it
decays (Thompson \& Duncan, 1995, 1996). See Woods \& Thompson (2004)
for a review of magnetar characteristics.

The variability of AXPs has for a long time been of interest both in the
X-ray band and the infrared for long-term and transitory events. For
example, whereas 1E~2259+586 and XTE~J1810$-$197 showed correlated
X-ray and infrared emission following outbursts (Tam et al. 2004;
Gotthelf at al. 2004), Durant \& van Kerkwijk (2005) found that for
1E~1048.1$-$5937 there was possibly an anti-correlation between the
infrared and X-ray fluxes. Nevertheless, the AXPs generally seem to
have consistent X-ray to infrared flux ratios.

4U~0142+61 was discovered at a 8.7s X-ray pulsar by Israel et
al. (1994), and is the brightest of the AXPs in the sky. It was
initially modeled with a hot black-body spectrum ($kT\approx
0.4$\,keV, e.g. White et al. 1996), and soft power-law at higher
energies. No narrow features have yet been found in the X-ray spectrum
in the 1--10\,keV range (Juett et al. 2002).

Hulleman et al. (2000, 2004) detected 4U~0142+61 in the optical and
infrared. The emission was found to be orders of magnitude below the
extrapolation of an X-ray power-law fitted to the X-ray spectrum, yet
orders of magnitude above the extrapolation of an X-ray black-body
component. By excluding the possibility of a faint binary companion or
of substantial accretion from supernova fall-back material, they
excluded two alternative scenarios to the magnetar model. The optical emission was
found to be pulsed at the pulsar period, with a large pulsed fraction
(Kern \& Martin, 2002; Dhillon et al. 2005), clearly identifying it as
magnetospheric in origin.

Recently, 4U~0142+61 has been detected in new spectral windows: Wang
et al. (2006) detected the object in two {\em Spitzer} imaging bands,
attributing the measured flux to thermal emission from a dusty
circumstellar disc; and den Hartog et al. (2006) identified the AXP
from long ($\sim$2MS) INTEGRAL observations - they found it has a
rising spectrum in the 20--150keV range which dominates the
energetics. 

Durant \& van Kerkwijk (2006) determined the interstellar extinction
to 4U~0142+61 by direct measurement of the optical depths in
individual photo-electric absorption edges from high-resolution X-ray
spectra. They found that the inferred column density was 40\% less
than had typically been stated from broad-band spectral modeling,
which solved a long-standing inconsistency between the implied
reddening to the source and the total reddening along the line of
sight (see Hulleman et al. 2004). This also revealed a possible, broad
spectral feature in the X-ray spectrum near 830\,eV (13\AA), a first for an
AXP. 

In this paper we present many different observations at different
epochs, from different observatories and in different parts of the
spectrum. Some of these we take from the literature, some we have
obtained from archival data and some are from our own observations. We
describe in detail the data and reduction for both the observations we
obtained and those retrieved from the archives. The following sections
enumerate the observations in decreasing order of wavelength. In
Section \S\ref{results} we compare both our data and those in the
literature and investigate the time-scales, correlations and spectral
dependence of the variability seen.

\section{Near-Infrared Observations}
Table \ref{IR} lists all the observations of 4U~0142+61 we are aware
of in the near-infrared (NIR), spanning several years. The Keck
magnitudes were taken from Hulleman et al. (2004), and we obtained new
observations from Gemini. For the other observations (from CFHT and
Subaru), we analyzed data from the archives. All the final magnitudes
and associated uncertainties are listed in Table \ref{IR}. We briefly
describe the reduction and analysis procedures we followed for each of
the observations that we analyzed.

For the K-band, we have analyzed observations taken in the three
similar, but not identical, filters: K, K$_S$, and K'. Each has a roughly
rectangular bandpass, centered on $\lambda_{eff}=2.2$, 2.15 and
2.12\,$\mu$m and with width 0.16, 0.15 and 0.18\,$\mu$m,
respectively. Using K$_S$ as our baseline, the overlap between the
filters is greater than 80\% for K and greater than 90\% for K'. From
the standard star list of Oersson et al. (1998) and their
interpolation to K'\footnote{see {\url
    http://www.mpia-hd.mpg.de/IRCAM/FAINTSTD/faintstd\_kprime.html}},
we find that the magnitude difference between K$_S$ and K, and K' and K
is between 0 and 0.04\,mag for a range of stellar colors. Since
4U 0142+61 has an intermediate $H-K$ color, the uncertainty on the K
flux from assuming that the filters are identical is at most
4\%.

\subsection{Gemini\label{gem}}
The Gemini observations were taken in a concerted effort to
investigate the infrared variability of 4U~0142+61, using NIRI, the
Near Infrared Imaging instrument (Hodapp et al. 2003) at
Gemini-North. NIRI is available for both imaging and spectroscopy, and
can be used as the detector for the output of ALTAIR, the Gemini
adaptive optics (AO) system. Here we do not use AO. The plate scale
is 0\farcs11 per pixel on the 1024 square Aladdin array. 

The observation are from four separate nights from September 2003 to
July 2005. On the 2, Nov 2004 H-and J-band images were obtained in
addition to the K-band images.  The last of our Gemini infrared
observations (July 2005) was taken under a DDT proposal to attempt to
observe 4U~0142+61 across the whole EM spectrum quasi-simultaneously
(see Den Hartog et al, in prep.).

We created final images after subtracting dark frames and dividing by
a flat-field image derived from the science images. For the $K_S$
image, we measured our 
magnitudes relative to the stars listed in Hulleman et al. (2004). We
find that the relative zero-point offsets accurate to $\approx0.016$\,mag, and
that the magnitudes measured for the AXP are 
in-between the $K$ and $K_S$ magnitudes presented by Hulleman et al. In November
2004, we also obtained images of standard stars FS-34, FS-112 and
FS-145 in all three filters, and used these to calibrate the magnitude
zero points of the J- and H-band images. To do this, we calculated an
aperture correction for the science images by performing aperture
photometry around our PSF stars with the same large aperture used to
measure the standard stars. These calibrations are not as accurate as
for the K-band (uncertainty $\approx0.025)$\,mag), but the detections
are poorer, so this does not contribute any additional uncertainty.

\subsection{CFHT}
Israel et al. (2004) reported observations of 4U~0142+61 on 18 August
2002 with the Adaptive Optics Bonette (AOB, Rigaut et al. 1998) of
Canada-France-Hawaii Telescope (CFHT), Hawaii.  The adaptive optics
system produces PSFs of reduced size in order to increase the
signal-to-noise ratio, and the corrected beam is imaged by KIR, a 1024
square Hawaii infrared detector with 0\farcs035 per pixel. We
retrieved the data from the CADC archive and re-analyzed them in order
to derive accurate relative magnitudes. The frames were
dark-subtracted and then flat-fielded using the median of the data
frames. The field of view is relatively small and contains only a
handful of stars. The FWHM of the PSF near the AXP was
$\approx0\farcs$14.

We performed photometry using the PSF-fitting package, {\tt daophot}
(Stetson, 1987). The PSF varies somewhat across the field of view, so
we constructed the model PSF from a few stars distributed across the
field. The analytic portion of the PSF was best fitted by a Lorentzian
function, and the residual image showed no systematic effects.

To calibrate the magnitude zero-points for the K-band image, we used
the photometry of several stars in the field listed by Hulleman et
al. (2004). The K$_S$ band and K' band do not exactly co-incide, but
they are very close in effective wavelength (2.15$\mu$m and 2.12$\mu$m
respectively), and so we treat them as identical, in the absence of
standard stars observed on the night. To calibrate the J- and H-band
photometric zero points, we used our images from Gemini (above).

We verified our calibration with two 2MASS stars on the image, which
gave consistent zero points. We chose to use the relative photometry as it
does not add any 
additional uncertainty. With the relative photometry we compare the
brightness of our source with stars of similar brightness rather than
the 2MASS stars which are much brighter.

\subsection{Subaru}
4U~0142+61 was imaged with the Infrared Camera and Spectrograph for
the Subaru Telescope (IRCS, Kobayashi et al. 2000) on two successive
nights, Sep 3 and 4, 2003 - see Morii et al. (2005). IRCS has a 1024
pixel square Aladdin array, with 0\farcs053 per pixel. On the first
night, only K-band imaging was performed, and on the second night J, H
and K. We retrieved the data from the archives (SMOKA) and
reduced them.

The IRCS detector suffers from a number of cosmetic artifacts, and
variable sensitivity across the detector. In order to successfully
perform the flat-fielding, we found the screen flats only partially
useful, leaving medium to large scale variations uncorrected. We
created flat-field images by taking the median of the science images,
scaled by the mode of their data and after rejecting outliers. For the
K-band images of the second night (showing the 
largest variations across the field and with a variable sky
background), we used both the screen flat, 
followed by the median of the partially corrected images.

The photometry for the $K_S$-band was calibrated relative to  Hulleman
et al. (2004) and relative to the Gemini frames above for the H- and
J-bands, as for the CFHT images. The 2MASS stars
in the field again give consistent zero-point offsets to our method.

\begin{deluxetable}{lccccc}
\tablecaption{Infrared observations of 4U~0142+61\label{IR}}
\tablewidth{0pt}
\tablehead{ \colhead{Date} & \colhead{MJD} &
  \colhead{Telescope/Instrument} & \colhead{$K$\tablenotemark{a}} &
  \colhead{$H$} & \colhead{$J$} }
\startdata
1999-02-08 & 51393 & Keck-I/NIRC & $K=19.68\pm0.02$\tablenotemark{b} \\
2001-10-30 & 52213 & Keck-I/NIRC & $K_S=20.15\pm0.08$\tablenotemark{b} \\
2002-08-18 & 52505 & CFHT/AOB & $K'=19.76\pm0.05$ & $20.52\pm0.11$ & $21.96\pm0.12$\\
2003-09-08 & 52891 & Subaru/IRCS&$K'=20.18\pm0.08$& \\
2003-09-09 & 52892 & Subaru/IRCS&$K'=20.78\pm0.08$& $20.90\pm0.08$ & $22.18\pm0.09$\\
2003-09-14 & 52897 & Gemini/NIRI&$K_S=19.85\pm0.04$\\
2003-10-29 & 52942 & Gemini/NIRI&$K_S=19.83\pm0.03$\\
2004-11-02 & 53312 & Gemini/NIRI&$K_S=19.96\pm0.07$&$20.69\pm0.12$&$21.97\pm0.16$\\
2005-07-26 & 53578 & Gemini/NIRI&$K_S=19.96\pm0.10$
\enddata
\tablecomments{Uncertainties are 1-sigma, and do not include
  zero-point errors (which are small in comparison); they are accurate relative to
  one-another.} 
\tablenotetext{a}{Magnitudes refer to the K, K$_S$ and K' bands, as shown. These
  are all approximately interchangeable (see text).}
\tablenotetext{b}{Taken from Hulleman et al. (2004).}
\end{deluxetable}

\section{Optical Observations}
Table \ref{Optical} lists all the observations of 4U~0142+61 in the
optical over the last few years. We include both measurements made by
us from our own observations, and several from the literature. We do
not have access to any of the archival data, so re-analysis for better
relative photometry was not possible.

\begin{deluxetable}{lccccc}
\tablecaption{Optical observations of 4U~0142+61\label{Optical}}
\tablewidth{0pt}
\tablehead{ \colhead{Date} & \colhead{MJD} &
  \colhead{Telescope/Instrument} & \colhead{$I$} & \colhead{$R$} & \colhead{$V$}
}
\startdata
1994-10-31 & 49657 & Keck-I/LRIS & & $24.89\pm0.08$\tablenotemark{a} &
$25.62\pm0.11$\tablenotemark{a} \\
1999-09-06 & 51428 & Keck-II/LRIS &$23.84\pm0.06$\tablenotemark{a}
&$24.89\pm0.07$\tablenotemark{a} \\
2002-09-12 & 52530 & WHT/ULTRACAM &$i'=23.7\pm0.1$\tablenotemark{a}&
&$g'=27.2\pm0.2$\tablenotemark{a} \\
 & & & $23.9\pm0.2$\\
2003-01-02 & 52642 & Keck-II/ESI & $23.77\pm0.11$ & $25.29\pm0.19$ & $26.10\pm0.18$\\
2003-09-04 & 52887 & UH88/OPTIC & $23.97\pm0.09$\tablenotemark{a} &
$25.58\pm0.18$\tablenotemark{a} & $25.32\pm0.13$\tablenotemark{a} \\
2003-12-21 & 52995 & Keck-II/ESI & $23.44\pm0.18$ & $25.34\pm0.12$\\
2005-07-13 & 53566 & Gemini/GMOS & & $r'=25.42\pm0.06$\\
 & & & & $25.2\pm0.2$
\enddata
\tablecomments{Magnitudes that are not in the standard Johnson-Cousins system
  are listed under the nearest band, and the appropriate filter is
  given. These are based on the Sloan filter-set and have
magnitudes which follow the AB system, rather than Vega
magnitudes. The estimated Johnson-Cousins magnitudes are given
immediately below each (for $g'$ this is not possible, since it lies
blueward of V).}
\tablenotetext{a}{Taken from the literature: Hulleman et al. (2004),
  Dhillon et al. (2005) and Morii et al. (2005)}
\end{deluxetable}

\subsection{Keck}
We obtained two nights of observations of 4U~0142+61 using the
Echellette Spectrograph and Imager (ESI; Epps \& Miller, 1998) at
Keck-II, Hawaii. In 
imaging mode, the instrument provides a standard set of
Johnson-Cousins filters, with the exception of the R-band, where the
Ellis filter is used ($R_E$), which is slightly shifted from the
standard Cousins R filter.

The detector employs dual-amplifier readout, and the bias has to be
subtracted from the two separate regions of the images (the bias is
easily derived from over-scan regions). In the case that the field of
view was located wholly within one amplifier area, we discarded the
other half of the image. We created flat-field images from
screen-flats taken on each night, and registered and stacked the
images after the correction. Since the observations in each band
typically consisted of only two images, some artifacts and cosmic rays
remain on the final image. None of these fall close to the object of
interest, and they do no affect our measurements.

To calibrate the photometric zero points, we compared our instrumental
magnitudes (from PSF fitting) with those listed in Hulleman et
al. (2004). Whereas the I- and V-bands required no color term, the
$R_E$-band does, since it does not well match the standard
Cousins R-band. Fortunately, we identified enough stars in the December
observation to achieve a well-defined zero-point offset versus $m_r-m_i$
color. For the January 2003 observation, rather that repeat this
process, we found the relative offset from one $R_E$ image to the
other and used the same calibration (since the January observation was
less good). Photon-noise dominates the uncertainty in the
magnitudes listed in Table \ref{Optical}.

\subsection{Gemini}
As part of the simultaneous, multi-wavelength observation campaign for
4U~0142+61 (see \S\ref{gem} above), we obtained imaging data with
the Gemini Multi-Object Spectrograph (GMOS-North, Hook et
al. 2004). GMOS is equipped with an integral field unit, but we used it
in imaging mode, with one amplifier covering a 1024$\times$2304 pixel
region with a pixel scale of 0\farcs07.
Unfortunately, the instrument is equipped with the SLOAN
filter-set rather than Johnson-Cousins filters, so we opted for the
middle of the optical band with the r' filter.

We subtracted the bias and applied a screen-flat correction, before
stacking the images, and analyzing them with {\tt daophot} as
before. For the calibration, we used again the photometry listed in
Hulleman et al. (2004), interpolating between the R- and V-bands using
the relationship in Smith et al. (2002). Since we do not
have a measure of color from the GMOS observations, we cannot interpolate the
magnitude to either the R- or I-bands for comparison with earlier
observations. However, we can obtain $r'$ from earlier
measurements. Doing this, we find that the r' magnitude is
consistent with that inferred from the 01/2003 observations, but not from
the other pairs of V and R magnitudes.

\section{X-ray Observations}
4U~0142+61 has been the subject of a long-term monitoring campaign
with the Rossi X-ray Timing Explorer (RXTE) satellite. These
observations have yielded roughly weekly measurements 
of the pulsed flux and pulse period of the pulsar. In addition, there
have been extensive observations with XMM-Newton (the X-ray
Multi-mirror Mission, see below) and
Chandra (e.g. Juett et al. 2002). We do not consider the Chandra data, since
there have been too few observations and using different
instruments/modes to be useful.

\subsection{RXTE}
The Proportional Counter Array (PCA) instrument of RXTE has five
proportional counting units, and provides very
high time-resolution and moderate energy resolution in the 2--60\,keV
energy range (Jahode et al. 1996). The pulsed flux
measurements are much more reliable than the total flux due to the
non-imaging nature of the PCA
instruments, and the nearby variable high-mass binary X-ray pulsar
RX~J0146.9+6121 (Motch et al. 1991). Observations were
made for the epoch $MJD=51700$ to 53700.

The data was analyzed in the same procedure as described in Gavriil \&
Kaspi (2002): an average pulse profile was created from several
observations, and cross-correlated with each observation in order to
obtain the (barycentred) average pulse arrival times, without losing coherence
from one observation to the next. 

The timing of the pulsar was found to be stable across the whole
epoch, and well-described by a single ephemeris.
Phase residuals are less than 0.05 cycles for the entire time-span.

In the first analysis, it appeared that the pulsed flux increased
nearly linearly with time, by about 40\% across the whole
time-span. On closer inspection, this was found to be the case for
only one of the five PCUs, and hence its data were discarded. We here
use the pulsed flux measurements only from the remaining (four or less)
PCUs. See Dib et al. (2006) for details.

\subsection{XMM-Newton}
In addition to the RXTE data, we analyzed archival {\em XMM-Newton}
data, which is able to measure the total flux rather than just pulsed
flux due to its better spatial resolution. Four observations exist in 
the archives, and these too span the range of time comparable to the
other data. XMM has three telescopes and five instruments which take
data simultaneously, but we decided only to use the data from the Reflection
Grating Spectrometers (RGS; den Herder et al. 2001), since they were
used in the same observation mode for all four observations. 

We used the standard pipeline and the Science Analysis Software,
(6.1.0, 2004-11-22) and the latest calibration files. For the final
(background-subtracted) fluxes, we used the SAS task {\tt
  rgsfluxer}. Although this task is not recommended for detailed spectral
analysis, for calculating the total flux, its accuracy is ample.

In the range 515eV to 1550eV (8 to 15\AA, where the 
sensitivity and calibration are well known), we find statistically
significant variability between the observations, but only by about
10\% at most, see Table \ref{XMM}.

\begin{deluxetable}{lccccc}
\tablecaption{XMM-Newton/RGS soft X-ray observations of 4U~0142+61\label{XMM}}
\tablewidth{0pt}
\tablehead{ \colhead{Obs Code} & \colhead{Date} & \colhead{MJD} &
  \colhead{Flux (8--15\AA)}\\
& & & \colhead{($10^{-11}$ erg s$^{-1}$ cm$^{-2}$)}
}
\startdata
0112780301 & 2000-12-28 & 51907 & $2.91\pm0.04$\\
0112481101 & 2003-01-24 & 52664 & $3.29\pm0.03$\\
0206670101 & 2004-03-01 & 53066 & $3.192\pm0.013$\\
0206670201 & 2004-07-25 & 53212 & $3.229\pm0.017$
\enddata
\tablecomments{1-sigma errors are derived from the XMM task {\tt rgsfluxer}}
\end{deluxetable}

\section{Results\label{results}}
\begin{figure}
\includegraphics[width=\hsize]{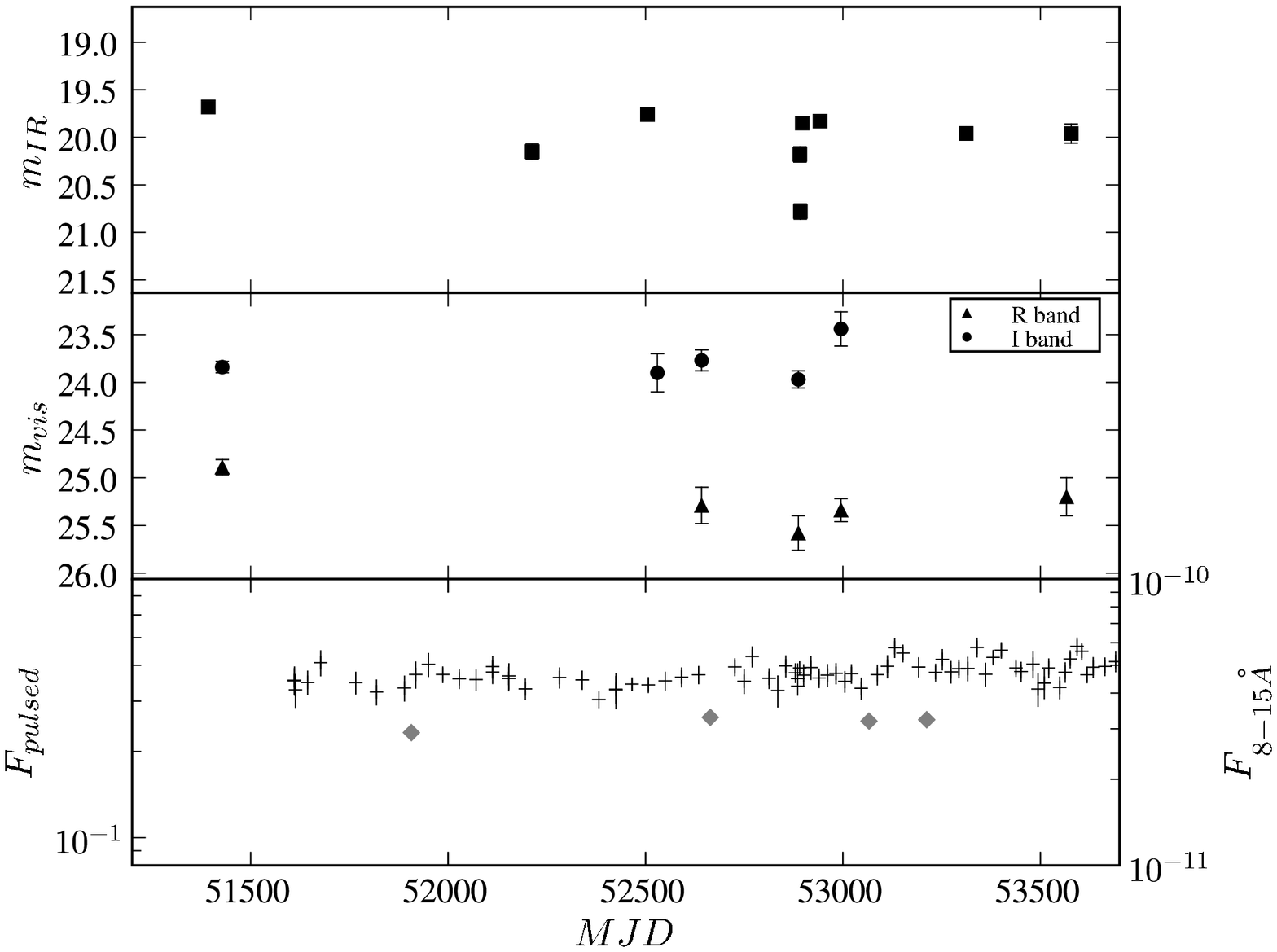}
\caption{Compilation of flux data from the infrared (K-band, top), optical
  (middle) and X-ray (bottom). In the latter we show both pulsed flux
  from RXTE (pluses, in (cts s$^{-1}$ PCU$^{-1}$)\,) and total flux
  from XMM/RGS (grey diamonds, right hand scale, in (erg s$^{-1}$
  cm$^{-2}$)\,). The X-ray fluxes are shown on logarithmic scales and
  span the same range as the optical/infrared, to
  match the logarithmic nature of the magnitude
  system.\label{bigplot}} 
\end{figure}

We plot the results of the measurements in infrared, optical and
X-rays in Figure \ref{bigplot}. For the optical magnitudes made under
the AB magnitude system (from WHT/ULTRACAM and Gemini/GMOS-N), we
estimate the appropriate Johnson-Cousins magnitude based on the
transformations of Smith et al. (2002). In neither of the cases do we
have $(r'-i')$ colors, but these can be estimated from the range of
$(R-I)$ colors from the other observations. We show the points with
increased error-bars to account for this uncertainty (see also Table
\ref{Optical}). 

\subsection{Variability Time-scales}
Whereas 4U 0142+61 was  regularly observed with RXTE
roughly once a week, the shortest delay between subsequent
measurements occurred in the infrared  
K-band, with three observations within a week in September 2003. We
find that the source varied by over a magnitude over a time-scale of
days. Such rapid and large variability was not expected and was not
seen before. Even more surprising, is the rapid dimming between the
two Subaru nights -- a rapid brightening could have been explained as
originating in some kind of energetic outburst. 

\begin{figure}
\includegraphics[width=0.49\hsize,angle=270]{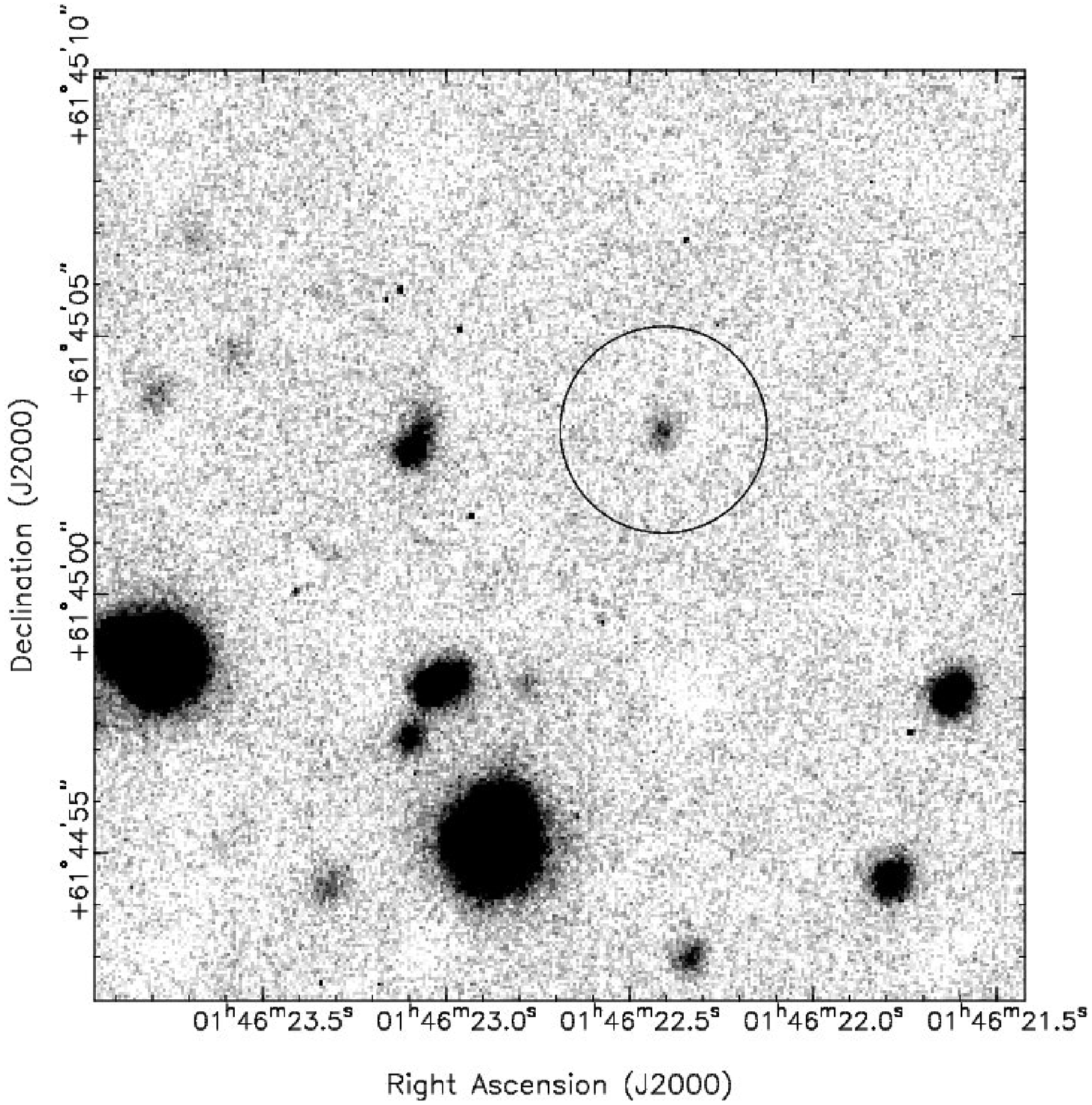}
\includegraphics[width=0.49\hsize,angle=270]{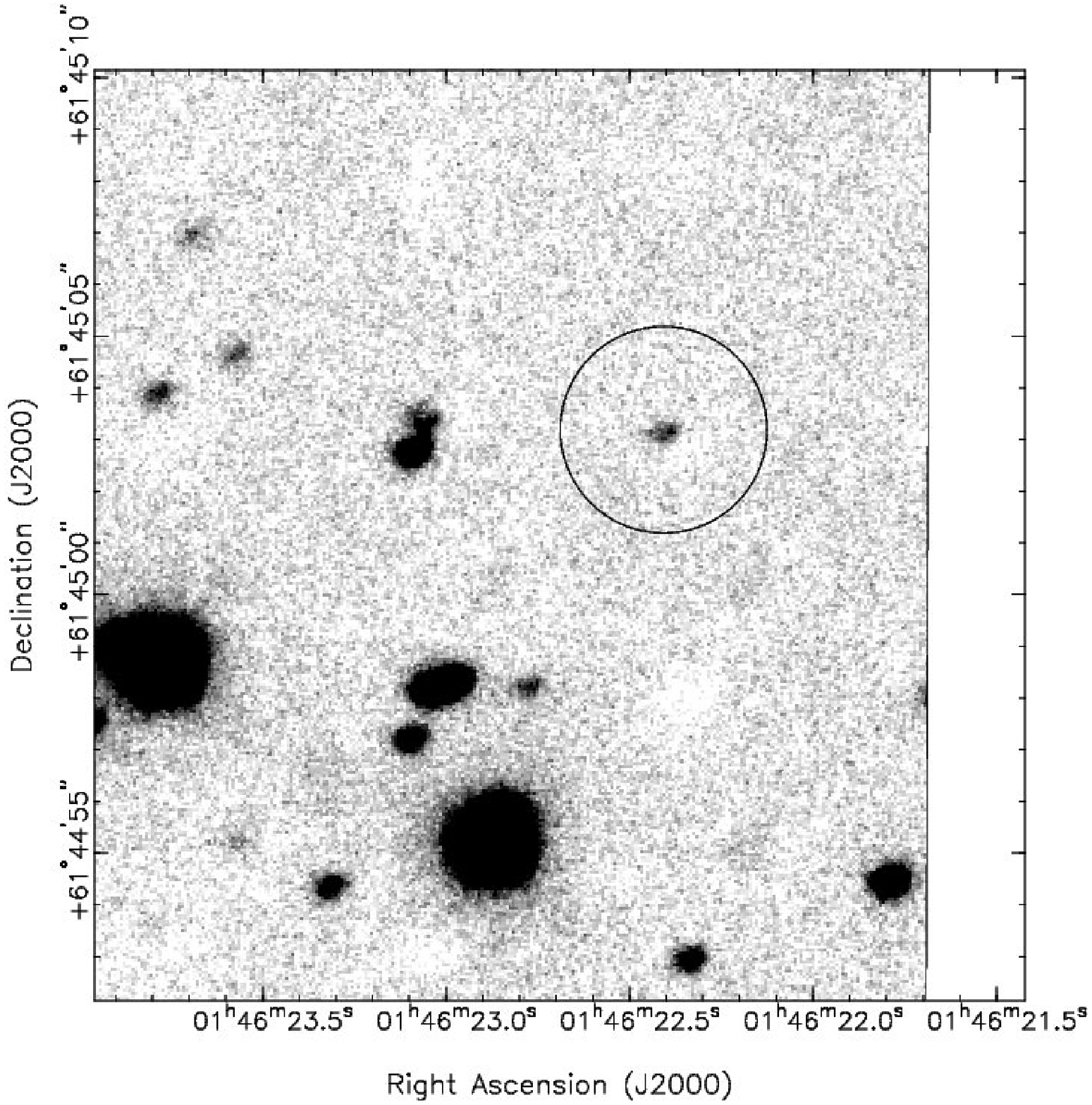}
\includegraphics[width=0.49\hsize,angle=270]{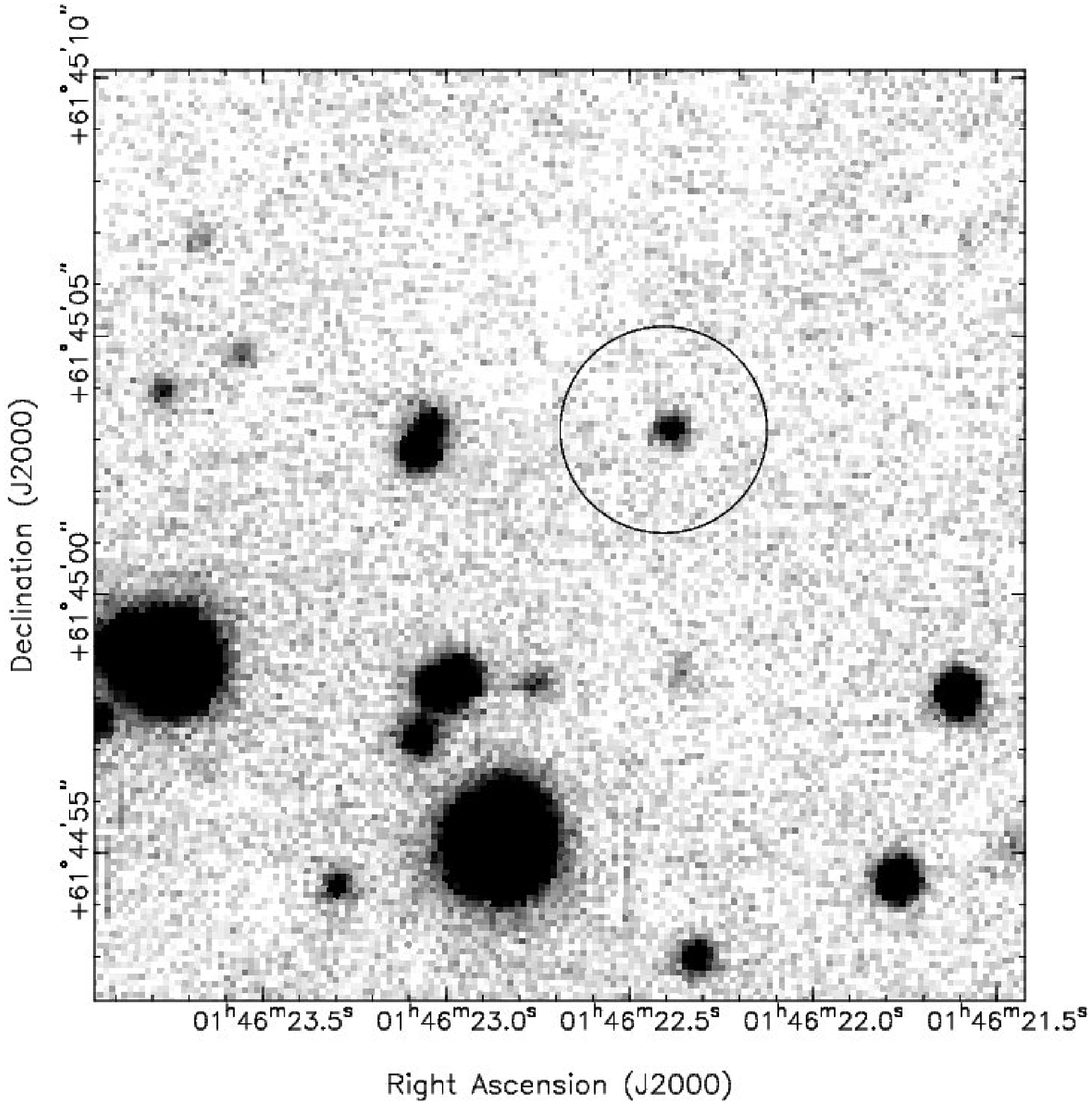}
\caption{K-band images of 4U~0142+61. Top left is the Subaru image
  from 2003-09-08 ($K'=20.18$), top right is the Subaru image from
  2003-09-09 ($K'=20.78$) and bottom is the Gemini image from
  2003-09-14 ($K_S=19.85$).
 The object is in the  center of each over-drawn circle, and 
  clearly varies compared to field stars}\label{images}
\end{figure}

In figure \ref{images} we show the three images of 4U~0142+61 in
question, and clearly the object does vary substantially. The
background blemish visible in the Subaru images near the source is
outside of the PSF fitting radius and does not affect the photometry. 

In the other wave-bands, such rapid variability is not seen, but the
sampling has not been dense enough for us to be able to discount the
possibility. Only in the X-ray can we state that for the longest
observation, 44.1ks for the March 2004 observation, the flux remained
constant within the observation to high precision (in the
8--15\AA\ range). 

\subsection{Correlations}
\begin{figure}
\includegraphics[width=\hsize]{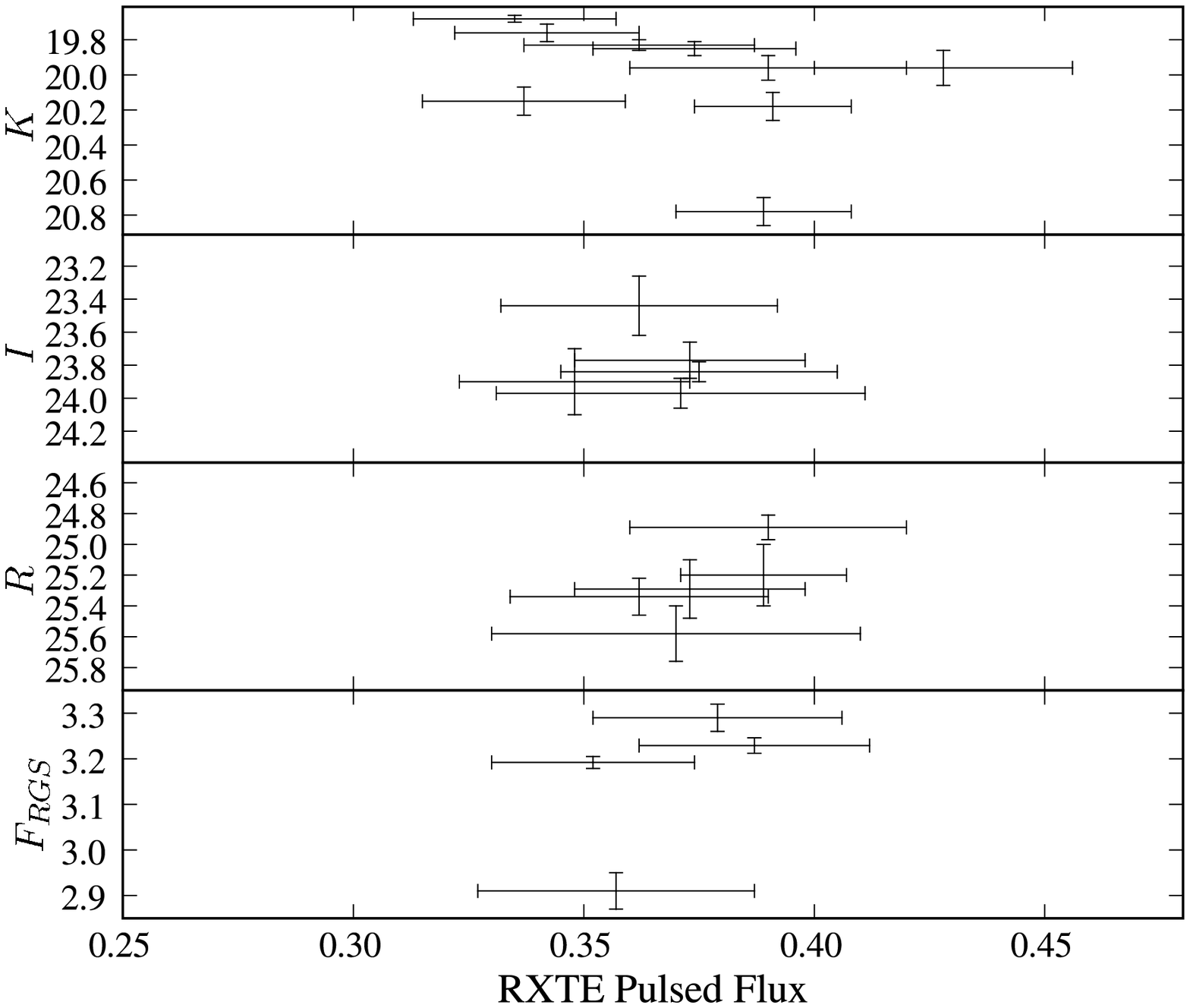}
\caption{Graphs of the various measures of flux versus the only
  time-series which was continuous across the whole epoch: the RXTE
  pulsed flux (2--10\,keV, in units of cts/s/PCU). RGS flux is in the
  8--15\AA\ range, in units (10$^{-11}$\,erg s$^{-1}$cm$^{-2}$). \label{corrplot}} 
\end{figure}

We plot the flux measurement of 4U~0142+61 in the infrared, optical
and X-rays versus time in Figure \ref{bigplot}, and in Figure
\ref{corrplot} we plot the measures of flux against one-another. We
have calculated the RXTE pulsed flux appropriate for each observation
in another band by linear interpolation between
the two closest measurements in time. 

No clear picture emerges from the two Figures. There is no apparent
correlation between the K-band magnitude and pulsed 
flux. For the I and R bands and the XMM/RGS flux, the range of values is not
big enough compared to the uncertainties to be able to make a definite
statement.

Intriguingly, the large feature in the time-series of K-band
photometry starting around $MJD=52892$, showing a rapid dimming and
re-brightening, has no change in the pulsed X-ray emission at that
time. At this time, the source was at its faintest in $I$ and $R$
(but not $V$).

\subsection{Spectral Changes and Components}
\begin{figure}
\includegraphics[width=\hsize]{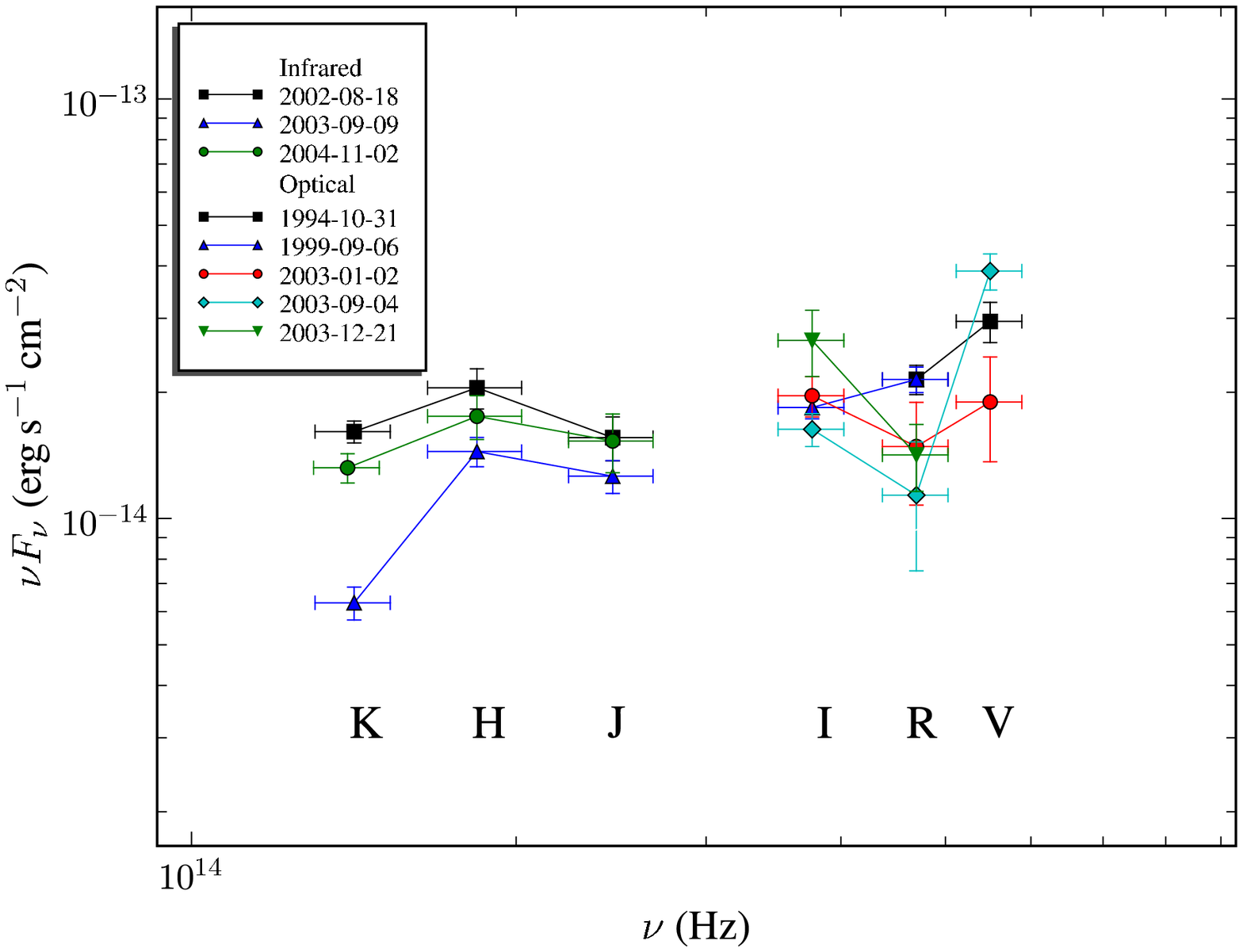}
\caption{Variability in the spectral energy distribution of 4U 0142+61
  for the different observations in the optical and infrared. We omit
  the ULTRACAM measurements, since the uncertainties are large. The
  points have been de-extincted using $A_V=3.5$ (Durant \& van
  Kerkwijk, 2006).\label{colors}} 
\end{figure}

In Figure \ref{colors}, we show infrared and optical spectral energy
distributions for each observations (each spectrum from within a
simgle night's observation).  In the infrared, the magnitude in H is
clearly correlated with K, but K varies
more than H. For J, it is not clear from our data whether it is also
related to K, or whether it is better described as remaining constant;
both are consistent with the data.

For the optical, the data show no apparent correlation at
all between $V$ and $R$. With only two simultaneous $V$ and $I$
measurements, we can say 
nothing of their possible relationship, except that both do vary (the
estimated I and V magnitudes from the ULTRACAM data have very large
uncertainties associated with them). The spectrum appears to be
markedly different for each observation.

\section{Conclusions and Discussion}
We have shown that the infrared emission of 4U~0142+61 varies by over
a magnitude on the time-scale of days, that there are no obvious
correlations between the infrared, optical and X-ray fluxes. The lack
of correlations suggests that
separate emission component are required for the various parts of the
spectrum. 

The first important point is that although we see no obvious
correlations, if the flux in any given wave-band is varying as fast as
it varies in
the infrared, the observations are not near enough to simultaneous to
be certain. Only a concerted campaign of simultaneous
observations at each wavelength would resolve this issue. In this
study we are not sensitive to shorter time-scales, so even faster
variability could in principle be occurring. On the other hand, the
longest XMM observations reveals a steady light-curve for over 44ks,
and significant changes are not seen in the individual exposures which
make up each optical/infrared observation. (The latter is not very
conclusive, however, as the signal in each frame is only small.)

The variability in the optical comes as a surprise, following Hulleman
et al's (2000) statement that they saw stability to within
0.02\,mag, and Dhillon et al.'s (2004) rough agreement with these magnitudes.
The variability is most pronounced in the R-band, and
suggests that perhaps there were transitory absorption
features (e.g. in the UH88 R-band observation). Beloborodov \&
Thomson (2006) suggest two likely mechanisms 
for the optical emission from a magnetar.
One possibility is from ions from the outer
magnetosphere, which absorb surface radio and microwave radiation at
their cyclotron resonance, and re-emit as they head nearer the poles
to higher cyclotron energies. A second possibility is coherent
curvature radiation from bunched pairs. Both mechanisms can in
principle explain the B-band cut-off seen by Hulleman et al. (2004),
but both also predict a smooth spectrum at longer wavelengths. These
predictions are valid, however, only for the equilibrium state, and
how they would be affected by variations of magnetic field and particle
kinetic energies is as yet unclear.

Wang et al. (2006) detected 4U~0142+61 in the mid-infrared using {\em
  Spitzer}, with fluxes well described as a cool thermal spectrum. They
proposed that the K-band emission also arises mainly from a dusty
circumstellar fall-back disc at the sublimation radius, which
reprocesses incident X-rays. If this were so, one would expect a
strong correlation with X-ray flux (with perhaps a time-lag). We see,
however, that the X-ray flux varies by much less than the K-band, that
there is no variability in the longest (44ks) XMM observation, and
that the pulsed component is roughly stable across the whole
epoch. In particular, we see no counterpart in X-rays to the dimming
even seen in the K-band.

These measurements relate, however, to only the soft
part of the X-ray spectrum, whereas den Hartog et al. (2006; see also
Kuiper et al. 2006) have
shown from INTEGRAL data that a rising power-law component in the
range 20--150keV dominates the energetics. Due to the low photon-flux
and instrumental sensitivity, time resolution of the order of days is
not currently possible, and it is unclear whether any variability has
been seen in the INTEGRAL observations to date (P. den Hartog, 2006,
pers. comm.).

If the infrared emission is magnetospheric in origin (e.g. from
cyclotron emission), one would expect it to be varying on the fastest
time-scales. This is because the region with cyclotron energies in the
infrared range is farthest from the neutron star and thus contains the
smallest inertia both in particles and in the magnetic field energy (the
latter being dominant). We have shown from the correlation between the
infrared magnitudes in KHJ that the component responsible for the
variability in K affects also H and possibly J, but is dominant
towards the long-wavelength end of the infrared.

The lack of obvious correlations between the various spectral bands
comes as another surprise, compared to cases such as that of 1E~2259+586,
where the X-ray and infrared decreased on the same time-scale
following an outburst (Tam et al. 2004). The latter case may, however, have been
a special one, where the normal emission mechanisms were
overwhelmed by an extra reservoir of energy which was deposited at the
time of the flare. Either the spectral components are truly
independent and caused by unique emission mechanisms,  the
variability is of a larger scale and shorter time-scale than had been
previously thought, or there is some hysteresis in the system, which
creates a lag between the emission observed in different wave-bands.

It seems clear that in order to understand the variability of this
intriguing source, multi-wavelength observations must be made much
more frequently than done until now. Nevertheless, a number of hints and
interesting relationships seem to be revealing themselves through
persistent examination.

\acknowledgments 
\noindent{\bf Acknowledgments}
Thanks to Rim Dib and the people at McGill for the RXTE data. Thanks
to Shri Kulkarni and Derek Fox for the Keck observations. We made
use of data retrieved from the CFHT archive at CADC, from the Subaru
archive at SMOKA, and from the XMM archive at XSA.
 We acknowledge financial support from NSERC.

\end{document}